\documentclass{emulateapj}
\usepackage[english]{babel}

\usepackage{graphicx}
\usepackage{natbib}
\usepackage[varg]{txfonts}
\usepackage{epstopdf}
\usepackage{wasysym}

\usepackage{tabularx}
\newcolumntype{L}[1]{>{\raggedright\arraybackslash}p{#1}}
\newcolumntype{C}[1]{>{\centering\arraybackslash}p{#1}}

\shorttitle{\textit{Chandra} Observations of C/2012 S1 and C/2011 L4}
\shortauthors{Snios et al.}

\begin{document}

\title{\textit{Chandra} Observations of Comets C/2012 S1 (ISON) and C/2011 L4 (PanSTARRS)}
\author{Bradford Snios$^{1}$, Vasili Kharchenko$^{1}$, Carey M. Lisse$^{2}$, Scott J. Wolk$^{3}$, 
		Konrad Dennerl$^{4}$, and Michael R. Combi$^{5}$}
\affil{${}^{1}$ Department of Physics, University of Connecticut, Storrs, CT 06269, USA}
\affil{${}^{2}$ Planetary Exploration Group, Space Department, 
	Johns Hopkins University Applied Physics Laboratory, Laurel, MD 20723, USA}
\affil{${}^{3}$ Chandra X-Ray Observatory Center, 
	Harvard-Smithsonian Center for Astrophysics, Cambridge, MA 02138, USA}
\affil{${}^{4}$ Max-Planck-Institut f{\"u}r extraterrestrische Physik, D-85748 Garching, Germany}
\affil{${}^{5}$ Department of Climate and Space Sciences and Engineering, 
	University of Michigan, Ann Arbor, MI 48109, USA}

\begin{abstract}
\noindent We present our results on the \textit{Chandra X-ray Observatory} Advanced CCD Imaging Spectrometer (ACIS) observations of the bright Oort Cloud comets C/2012 S1 (ISON) and C/2011 L4 (PanSTARRS). ISON was observed between 2013 October 31--November 06 during variable speed solar wind (SW), and PanSTARRS was observed between 2013 April 17--23 during fast SW. ISON produced an extended parabolic X-ray morphology consistent with a collisionally thick coma, while PanSTARRS demonstrated only a diffuse X-ray-emitting region. We consider these emissions to be from charge exchange (CX) and model each comet's emission spectrum from first principles accordingly. Our model agrees with the observational spectra and also generates composition ratios for heavy, highly charged SW ions interacting with the cometary atmosphere. We compare our derived SW ion compositions to observational data and find a strong agreement between them. These results further demonstrate the utility of CX emissions as a remote diagnostics tool of both astrophysical plasma interaction and SW composition. In addition, we observe potential soft X-ray emissions via ACIS around 0.2 keV from both comets that are correlated in intensity to the hard X-ray emissions between 0.4--1.0 keV. We fit our CX model to these emissions, but our lack of a unique solution at low energies makes it impossible to conclude if they are cometary CX in origin. We lastly discuss probable emission mechanism sources for the soft X-rays and explore new opportunities these findings present in understanding cometary emission processes via \textit{Chandra}. 

\end{abstract}


\maketitle


\section{\label{Intro}Introduction}

Cometary X-ray emissions, originally discovered by \cite{Lisse1996} and now observed in over 30 comets, are a well-studied phenomenon. It has been shown that the majority of these emissions are caused by solar wind Charge Exchange (CX) interactions between highly charged, heavy solar wind (SW) ions ($\sim$0.1\% of all solar wind ions) and neutral gas ejected from the comet nucleus into the coma \citep{Cravens1997, Krasnopolsky1997, Kharchenko2003, Lisse2004b, Bodewits2007, Dennerl2010}. A simplified theoretical description of interaction between the SW plasma and cometary atmosphere shows that the emission originates predominantly from the sunward hemisphere of the neutral coma and creates a projected paraboloid of emission with the comet at its focal point \citep{Haberli1997,Wegmann2004}.  

Two recently discovered comets that were observed by the \textit{Chandra X-ray Observatory} are comets C/2012 S1 (ISON) and C/2011 L4 (PanSTARRS), both of Oort cloud origin. ISON was a comet first detected at $\sim$10 AU from the Sun and was well studied during its first close inner system perihelion passage. We observed the comet at moderate activity ($Q_{\textrm{gas}} \approx 10^{28}$ mol s$^{-1}$). ISON had also begun suffering from a series of fragmentation events near the end of our observations that markedly ramped up its outgassing activity \citep{Combi2014b}. Our observations were also taken at a time of variable SW speeds, as indicated by the \textit{Advanced Composition Explorer} (\textit{ACE}).

PanSTARRS, although a naked eye object from Earth in mid-March, did not have such a favorable close passage by the inner planets and was only sparsely observed. It did, however, demonstrate a fantastically rich outpouring of dusty material in 2013 March--April as it passed through perihelion, as seen by \textit{STEREO} \citep{Raouafi2015}. PanSTARRS is an unusually dust-rich comet, with a dust-to-gas mass ratio greater than 4 \citep{Yang2014}. By contrast, ISON was seen to be a dust-poor comet  with a dust-to-gas mass ratio less than 1 \citep{Meech2013}. Since the \textit{Chandra} observations for these two comets have not previously been analyzed, we decide to examine their emissions for detailed analysis and interpretation of the cometary X-ray emission spectra via modeling. 

\begin{table*}[t]
	\caption{\label{Parameter} \textit{Chandra} Comet Observation Parameters}
	\resizebox{\textwidth}{!}{
	\begin{tabular}{ l c c c c c c c c c}
		\tableline \tableline
				& & & $T_{\textrm{exp}}$ & $r_{\textrm{c}}$ & $\Delta$ & Lat$_{\astrosun}$ 
				& Long$_{\astrosun}$ 
					& $Q_{\textrm{H$_{2}$O}}$ & $v_{\textrm{p}}$ \\
				Comet & Obs. Date &  Prop. Num. & (ks) & (AU) & (AU) & (deg) 
					& (deg) & (10$^{28}$ mol s$^{-1}$) & (km s$^{-1}$) \\
		\tableline
		PanSTARRS & 2013 Apr 17--23 & 14108442 & 45 & 1.10 
			& 1.44 & 84.16 & 150.5 & 5\tablenote{\cite{Combi2014a}} & 377$^{*}$ \\
		ISON & 2013 Oct 31--Nov 6 & 15100583 & 36 & 1.18 & 0.95 
			& 1.130 & 115.0 & 2\tablenote{\cite{Combi2014b}} & 313 \\
		\tableline
	\end{tabular}}
	\tablecomments{Observation parameters are listed as follows: 
			\textit{Chandra} observation date, observation proposal number, 
			exposure time $T_{\textrm{exp}}$, comet-Sun distance $r_{\textrm{c}}$, comet-Earth 
			distance $\Delta$, Heliospheric Latitude Lat$_{\astrosun}$ 
			and Longitude Long$_{\astrosun}$, H$_{2}$O production rate $Q_{\textrm{H$_{2}$O}}$, 
			and solar wind proton velocity $v_{\textrm{p}}$ from the \textit{ACE-SWEPAM}
			online data archive. Due to the large difference in heliospheric latitude 
			between \textit{ACE} and PanSTARRS, it is unlikely they experienced similar SW 
			speeds. More likely, PanSTARRS encountered fast SW due to its 
			high altitude. We denote our uncertainty in the observed SW speed value 
			with an asterisk.}
\end{table*}

\begin{figure*}[t]
	\epsscale{1.17}
	\plottwo{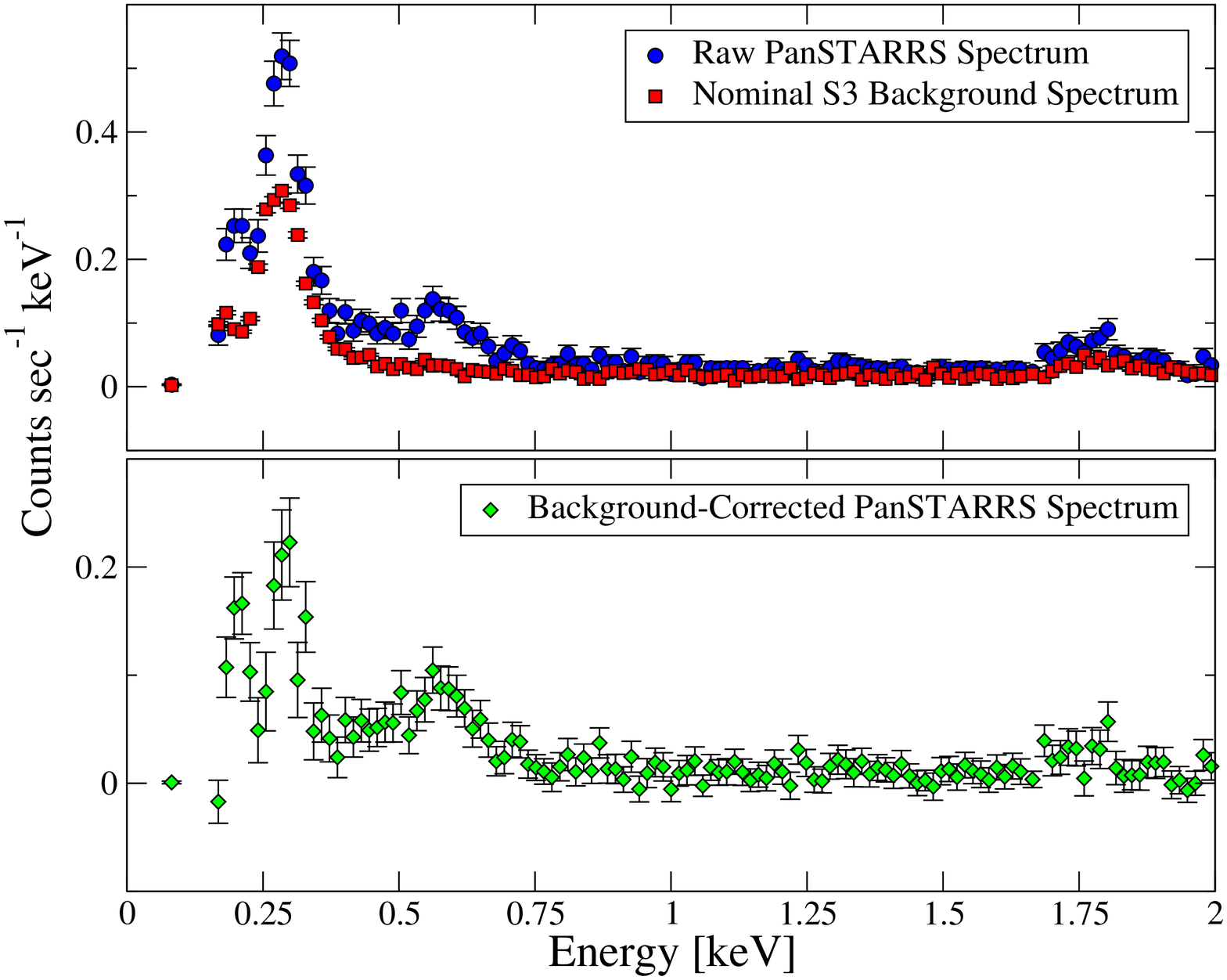}{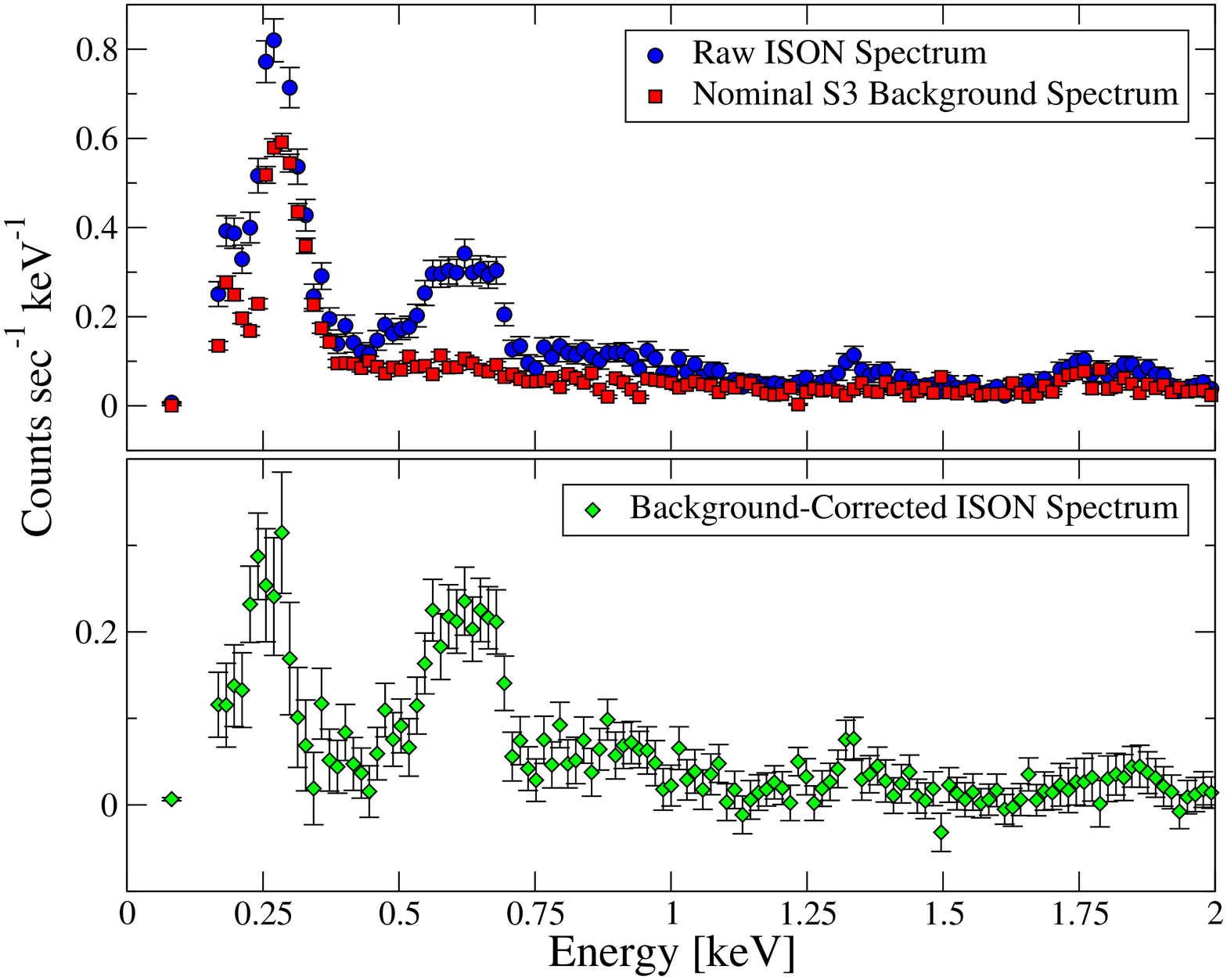}
	\caption{Summed total spectra of X-ray photon counts for comets PanSTARRS and ISON extracted from \textit{Chandra} observations. Both spectra utilize nominal S3 chip background emissions for background correction calculations.\\}
	\label{comets}
\end{figure*}

To model CX X-ray emissions for these comets, we decide to expand upon previous modeling techniques \citep{Kharchenko2000, Kharchenko2001, Krasnopolsky2002, Bodewits2007}. Though these models are robust, most only incorporate the primary emission lines, generally 10--20 lines, out of the possible 700+ lines that may be generated in an average CX interaction with a cometary atmosphere \citep{Kharchenko2003}. This is typically performed because each emission line is treated as a free modeling fit parameter, and increasing the total number of parameters will significantly reduce confidence in any results due to chi-square testing. However, proper consideration of state selective CX cross sections of highly charged SW ions will reduce model fit parameters as all emission lines per ion will be set at fix ratios determined by their cross sections and photon emission yields \citep{Bodewits2007}. Such a model would therefore only be dependent on the heavy SW ion composition, reducing the model from 700+ parameters down to 10-20. We therefore choose to develop a CX model from first principles that will include all possible lines arising in radiative cascading processes of excited SW ions with proper cross sections. This should simplify input parameters of cometary X-ray modeling through limiting of input variables to SW ion composition while also improving its physical accuracy through the increase of emission lines. Our model may also be utilized as a remote diagnostic tool for solar wind composition.

In this article, we analyze the \textit{Chandra X-ray Observatory} observations of comets C/2012 S1 (ISON) and C/2011 L4 (PanSTARRS). Each comet had unique conditions, either solar or cometary, that may impact CX emissions, and these different conditions should also provide an excellent test for our model. We describe details regarding the observations, data extraction, and our modeling techniques in Section 2. Our results are presented in Section 3. We discuss our findings in Section 4. Last, we provide a summary of our findings in Section 5.


\section{\label{Analysis}Observations and Analysis}

\subsection{Chandra Observations}

For both comets selected, the \textit{Chandra} observations were performed using the Advanced CCD Imaging Spectrometer (ACIS). The comet was centered on the S3 chip as it offers the most sensitive low-energy response in the ACIS array, and ACIS was set to very faint mode for all observations to assist in filtering out bad X-ray events, such as cosmic X-rays, from the source events. Each observation was also performed in drift-scan mode where no active guidance is enabled and \textit{Chandra's} pointing was only updated to re-center before the comet moved off the chip. 

SW proton velocities were extracted from \textit{ACE}, a satellite located at the L1 Lagrangian point that continuously records SW conditions. SW speed at each comet was calculated through time of flight corrections between \textit{ACE} and the comet observations, and the resulting SW velocities were found to be consistent with slow SW. However, we note the large discrepancy in heliospheric latitude between PanSTARRS and \textit{ACE} during our observations. Since PanSTARRS was at high latitude, we infer that it was bombarded with fast SW \citep{Geiss1995,Schwadron2000}. ISON was observed at similar heliospheric latitude to \textit{ACE}, so SW conditions should be similar between the two. In addition, solar X-ray activity detected by the \textit{GOES} X-ray satellite indicates that several M-class solar flare events occurred during ISON's observations, while solar activity was average for PanSTARRS. See Table \ref{Parameter} for additional details on the observation parameters for both comets. 

Since all observations were performed in drift-scan mode, we first convert all images to object-centered coordinates through use of the \textit{sso\_freeze} routine found in the \textit{Chandra} Interactive Analysis of Observations (CIAO) software package \citep{Fruscione2006}. We then generate our resulting spectra via CIAO's \textit{specextract} routine and are combined with the \textit{combine\_spectra} routine. All steps are performed with CIAO v4.7. The cumulative cometary and background X-ray emission spectra are shown in Figure \ref{comets}. 

\begin{figure}[t]
	\epsscale{1.10}
	\plotone{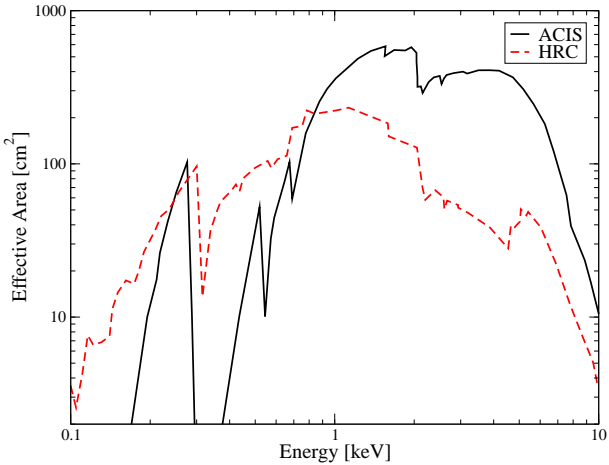} 
	\caption{A comparison of effective area functions for the ACIS and HRC instruments, as 
	documented in the \textit{Chandra} handbook. ISON observations were performed with both 
	instruments as HRC has a higher sensitivity to soft X-ray emissions than ACIS. \\}
\label{effective_area}
\end{figure}

In addition to these observations, ISON was also observed with the High Resolution Camera (HRC) on Chandra between the ACIS visits, also in drift-scan mode. HRC observations of a comet had never been performed in conjunction with ACIS observations before. Such observations were proposed for ISON due to HRC's increased sensitivity to soft X-ray emissions over ACIS, see Figure \ref{effective_area}, as it makes the two instruments complementary to one another. The resulting images and our discussion of their implications are in Section 4.1.  

\subsection{Spectrum Analysis}

\subsubsection{Background Correction}

Given the low count rate and extended nature of cometary X-ray emissions, both proper background correction and exposure map calculation are crucial. Since the \textit{specextract} routine in CIAO correctly handles any differences in exposure created from reprojecting an event to the reference frame of the comet, we focus our investigation on two possible background correction techniques that we could employ: on-chip and blank-sky corrections. 

On-chip correction is the most frequently utilized technique in cometary analysis and is performed by isolating a region of pure background on the same chip that observed the source. A background spectrum, assumed to be constant over the entire area of the chip, is extracted from this region and subtracted from the raw source spectrum to create the source spectrum. Such a method is valuable for X-ray analysis due to the high variability of X-ray background over time as it ensures near-identical background to that found in the raw source spectrum. However, observations of extended sources leave little area on chip for a proper background region to be defined. In the case of our comet observations, the source occupies 70--90\% of the S3 chip, which reduces the possible background statistics we can gather and can increase the uncertainty in the resulting source spectrum.

The alternative method is blank-sky background correction, which is performed by matching the coordinates of our comet observation to a similar blank-sky image where only background emissions were observed. By scaling the blank-sky observation to the exposure map of our original observation, we can generate a synthetic background spectrum that can be utilized with our data. This technique is beneficial for the low statistical uncertainty it introduces to our resulting spectrum. In addition, blank-sky correction is often preferred for extended sources, such as comets, as on-chip background regions may be contaminated by the source. Despite these benefits, the high variability of the X-ray background may result in blank-sky correction introducing random uncertainty into our calculations that would not exist from the on-chip method. 

In our analysis of both techniques, we find that the spectral uncertainty introduced via on-chip correction only becomes significant at energies greater than  2 keV for ISON and 1 keV for PanSTARRS. Since we are focused on analyzing the CX emissions up to 1 keV from each comet, we choose the on-chip correction method to avoid introducing additional uncertainty due to the X-ray background variability. We uniquely select the background area for each observation to avoid contamination from other on-chip astronomical objects as they varied significantly in location between each observation due to the close comet-detector distance and increasing comet velocity. See Fig \ref{comets} for the resulting background spectra and the background-corrected source spectra. We note that the resulting spectra possess large error bars relative to the scatter spread of the data, possibly indicating an unknown source of systematic error. All software tools and data reduction techniques were therefore tested separately for such an issue, and no sources of systematic error were found in our analysis. 

\subsubsection{\label{modeling}CX Modeling of Cometary X-Ray Emissions} 

As discussed in Section 1, a primary goal of our work is to develop a CX model from first principles that can provide more accurate diagnostic of the SW plasma interacting with cometary gas. 

We begin by expanding upon the CX model outlined in \cite{Kharchenko2000}, \cite{Kharchenko2001}, \cite{Krasnopolsky2002}, and \cite{Bodewits2007}. The emitted intensity $I$ of the photon flux induced by CX collisions is defined as the total emission resulting from the interaction between $k$ species of cometary atoms/molecules and SW ions $l$, where $l$ is dependent on both the element and its charge, within the cometary atmosphere. We define it as an integral over the line of sight distance $s$ and the solid viewing angle $\Omega_{s}$, 

\begin{equation}
	\label{I_total}
	I(\hbar \omega_{j}) = \sum\limits_{k,l} \int{n_{k} n_{l} \sigma_{k,l} |\vec{v_{k}}-\vec{v_{l}}| 
		P_{k,l}^{(j)}(\hbar \omega_{j}) ds d\Omega_{s} } ,
\end{equation}

\noindent where $n_{k}$ is the cometary particle density, $n_{l}$ is the SW ion density at the comet, $\sigma_{k,l}$ is the charge transfer cross section for collisions between $k$ neutrals and $l$ ions, $v_{k}$ is the cometary particle velocity, $v_l$ is the local SW velocity, and $P_{k,l}^{(j)}$ is the photon yield for emissions with the energy $\hbar \omega_{j}$ in the collision between $k$ and $l$ species. The total yield of all X-ray photons is normalized to 
unity, where $\sum\limits_{j} P_{k,l}^{(j)}(\hbar \omega_{j}) = 1$, per each unique $k$ and $l$ in order to be valid on a per collisional basis.

For our equation's parameters, both $n_{k}$ and $v_{k}$ are found from observational data on the comets \citep{Combi2014a,Combi2014b}. The physical parameters $P_{k,l}^{(j)}$ and $\sigma_{k,l}$ are obtained from previous lab and theoretical research \citep{Dijkkamp1985, Janev1985, Johnson1985, Kelly1987, Suraud1991, Cann1992, Janev1995, Wiese1996, Kharchenko2003, Koutroumpa2006, Koutroumpa2009}. In regards to $n_{l}$ and $v_{l}$, we set them equal to average values taken from previous analyses of the SW plasma \citep{Bochsler2007}.

We initially find that our modeled spectral intensity does not accurately fit the observational data due to the high variance in SW conditions as a function of both time and solar longitude, causing inaccurate modeled values for $n_{l}$ and $v_{l}$. Since we lack any direct observations of these values at the comet, we therefore allow these parameters to vary within physical limits until the best agreement between our relative modeled intensity and the observational intensity over the 0.3--1.0 keV energy range is found.

\subsubsection{CX Model Composition}

The CX emission spectrum in our model is computed for two independent major groups of heavy SW ions:

\begin{enumerate}

\item(Group A): key heavy ions for ``3/4 keV" energy interval (C$^{5+}$, C$^{6+}$, N$^{5+}$, N$^{6+}$, N$^{7+}$, O$^{6+}$, O$^{7+}$, O$^{8+}$, Ne$^{8+}$,and Ne$^{9+}$). This group includes the CX emission spectra generated from collisions between cometary neutrals  (primarily H$_{2}$O) and H-like, He-like, and Li-like heavy SW ions. The CX spectra of these ions are reasonable constrained by lab and theoretical researches \citep{Dijkkamp1985, Kelly1987, Suraud1991, Wiese1996, Kharchenko2003, Koutroumpa2006, Koutroumpa2009, Chutjian2012}.

\item(Group B): heavier excited ions (Mg$^{q+}$, S$^{q+}$, Si$^{q+}$, and Fe$^{q+}$) that primarily contribute to the soft X-ray spectra (below 0.4 keV). The cross sections and relative intensity of different emission lines of the CX cascading spectra for these ions are less known than for the ions from Group A but are well estimated \citep{Harel1998,Simcic2010}. The energy position of spectral lines are well defined \citep{NIST2015}.

\end{enumerate}

The spectra of CX cascading photons for Groups A and B are computed independently and then unified into a synthetic spectrum that represents the most probable emissions up to 1 keV. Ion elemental and charge composition for all groups are treated as variable parameters that are initially set to average SW composition ratios \citep{Bochsler2007, Lepri2013}. The SW composition is then varied until the $\chi^2$ value is minimized. Due to Chandra's low sensitivity below 0.35 keV and the lack of accurate calibration near the carbon K-shell line at 0.284 keV, we find that varying several SW ions types that predominantly emit in this region produces no change to $\chi^2$. As a result, these SW ion types are left constant as average SW composition ratios. The initial SW ratios and our resulting ratios for both comet observations are shown in Table \ref{Ratio}.


\section{\label{Results}Results}

\begin{figure*}[t]
	\center{
			\epsscale{1.17}
			\plottwo{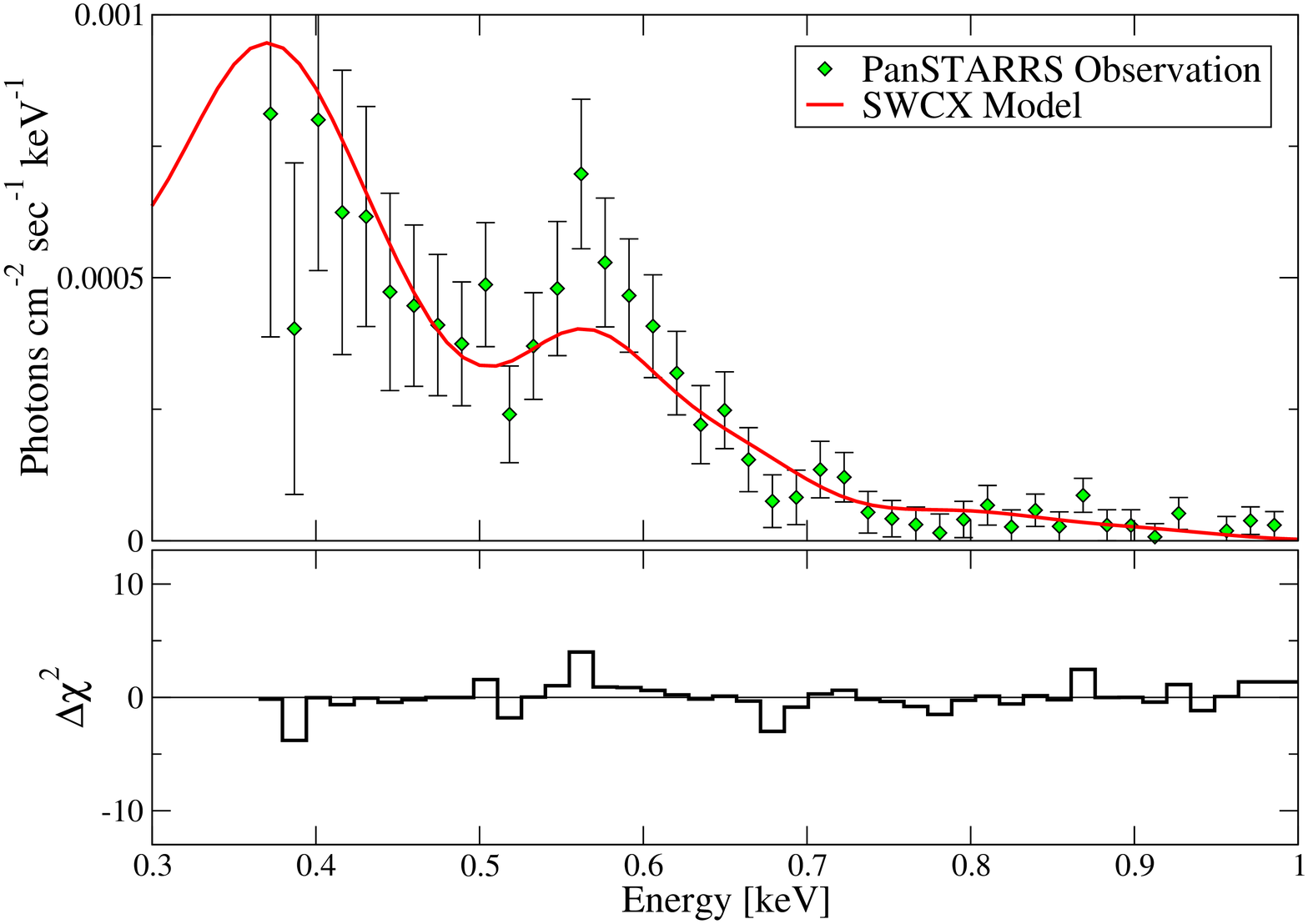} {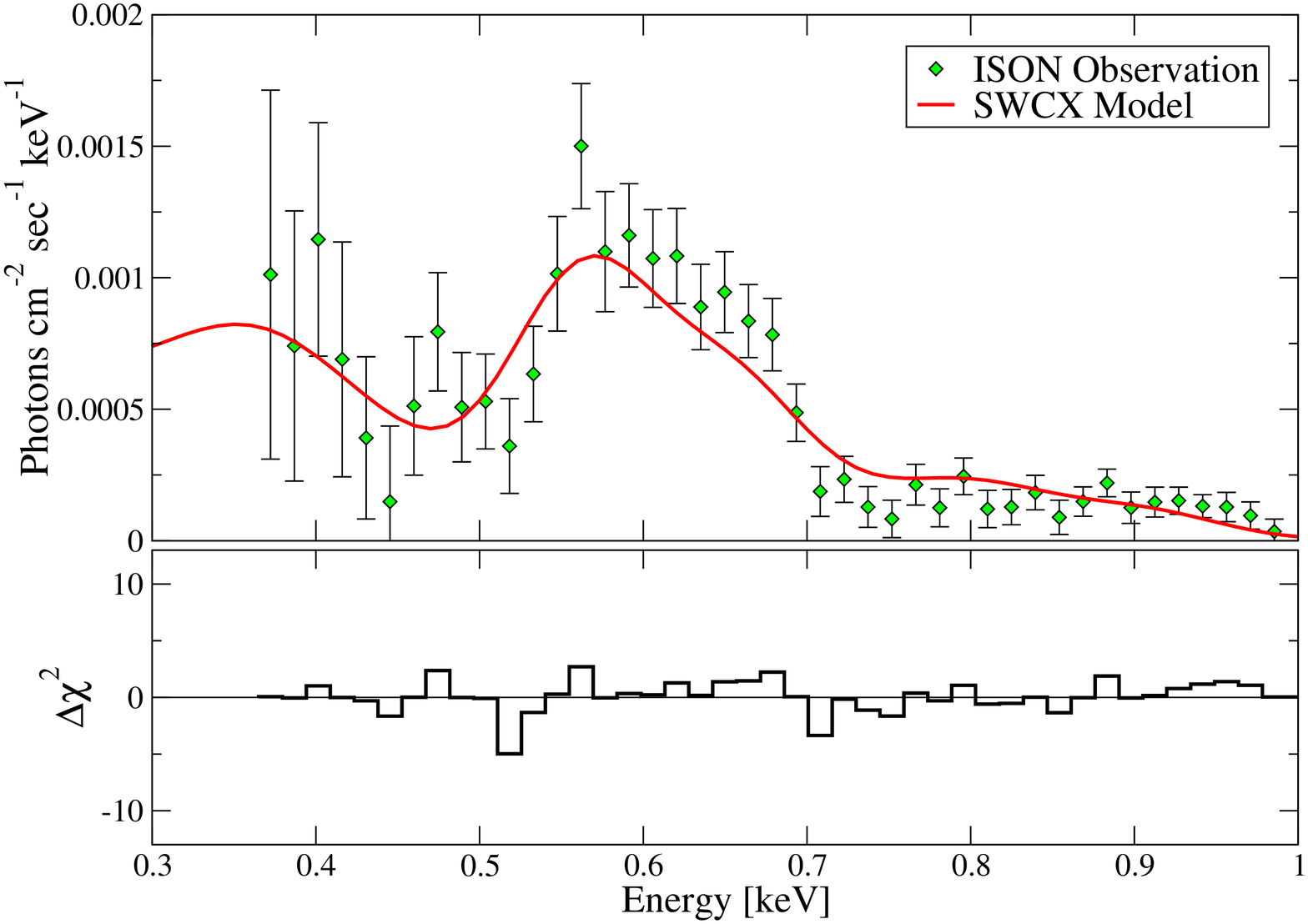}
	\caption{Comparison between our CX model and the average background-corrected observations 
	for comets PanSTARRS and ISON and the $\chi^2$ residuals ($\Delta \chi^2$) of the data-model
	comparison. Each observational spectrum has been grouped with a minimum of six counts per bin 
	for proper statistics. Our model varies SW composition ratios until a best-fit is found. 
	The resulting SW composition ratios for each model are detailed in Table \ref{Ratio}.\\}
	\label{swcx}}
\end{figure*}

Using the model outlined in the previous section, we create a theoretical CX spectrum for the two comets observed. Our theoretical spectra are compared to the average background-corrected observational spectra, and the results are shown in Figure \ref{swcx}. 

We also calculate the reduced $\chi^2$, also known as $\chi_{R}^2$, by dividing $\chi^2$ with the degrees of freedom (dof) for each comet. Both comet observations are binned with a minimum of six counts per spectral bin for proper gaussian statistics. For ISON, which has 32 dof, we find $\chi_{R}^2$ = 1.1 over the 0.35--1.00 keV range. For PanSTARRS, which has 30 dof, $\chi_{R}^2$ = 1.2 for the 0.35--1.00 keV range. All spectra are truncated at 0.35 keV due to the carbon K-shell absorption edge detector contamination present from ACIS at energies below this threshold. These results provide a more complete and physically accurate picture of the CX process in cometary atmospheres than found in the previous generation of models.  

\begin{table}[t]
	\centering
	\caption{\label{Ratio} SW Composition Ratio Inputs and Results}
	\begin{tabular}{l c c c | l c c c}
		\hline \hline
		 & Avg. & ISON & PS &  & Avg. & ISON & PS \\
		Ion & Ratio\tablenote{\cite{Bochsler2007}} & Ratio & Ratio 
		& Ion & Ratio$^{\textrm{a}}$ & Ratio & Ratio  \\ \hline 
			          
			C$^{6+}$  & 0.318 & 0.318 & 0.310 & Mg$^{10+}$ & 0.098 & 0.098 & 0.078 \\
			C$^{5+}$  & 0.210 & 0.240 & 0.240 & Mg$^{9+}$  & 0.052 & --- & --- \\ 
			N$^{7+}$  & 0.006 & --- & --- & Si$^{10+}$    & 0.021 & --- & --- \\ 
			N$^{6+}$  & 0.058 & --- & --- & S$^{11+}$   & 0.005 & --- & --- \\ 
			N$^{5+}$  & 0.065 & --- & --- & S$^{10+}$     & 0.016 & --- & --- \\
			O$^{8+}$  & 0.070 & 0.100 & 0.040 & S$^{9+}$     & 0.019 & --- & --- \\
			O$^{7+}$  & 0.200 & 0.200 & 0.100 & Fe$^{13+}$ & 0.002 & --- & --- \\
			O$^{6+}$  & 0.730 & 0.700 & 0.860 & Fe$^{12+}$ & 0.007 & --- & --- \\
			Ne$^{9+}$ & 0.004 & 0.020 & 0.004 & Fe$^{11+}$ & 0.023 & --- & --- \\
			Ne$^{8+}$ & 0.084 & 0.068 & 0.084 & Fe$^{10+}$ & 0.031 & --- & --- \\
		\hline
	\end{tabular}
	\tablecomments{Model-calculated SW ion ratios for comets ISON and PanSTARRS (PS) 
	in comparison to average slow SW ratios. All ratios are normalized with respect to the total 
	SW oxygen. All calculated values are found to have an average uncertainty of $\pm15\%$. 
	Values left blank are because the observational spectrum does not possess 
	the resolution required to accurately calculate those ratios, and so the model treats them 
	as constants.}
\end{table}

\begin{table}[t]
	\centering
	\caption{\label{ACE} SW Composition Comparison to \textit{ACE}}
	\begin{tabular}{l c c c}
		\tableline \tableline
		Source & C$^{6+}$/C$^{5+}$ & O$^{7+}$/O$^{6+}$ & O$^{8+}$/O$^{6+}$  \\
	  \tableline        
			ISON & 1.35 & 0.28 & 0.14 \\
			\textit{ACE} & $1.18{+0.80 \atop -0.48}$
				& $0.25{+0.12 \atop -0.08}$ & 0.09${+0.19 \atop -0.06}$\\
			\tableline
			PanSTARRS & 1.29 & 0.12 & 0.05 \\
			\textit{ACE} & $1.09{+0.62 \atop -0.39}$ 
				& $0.22{+0.13 \atop -0.08}$ & $0.08{+0.14 \atop -0.05}$\\
		\tableline
	\end{tabular}
	\tablecomments{A comparison between the model-calculated SW ion ratios and the average
			 values  observed by \textit{ACE}. All calculated values are found to have an average 
			 uncertainty of $\pm20\%$ and agree to the observational 
			 data within uncertainty. We note that agreement between \textit{ACE} 
			 and PanSTARRS is inconclusive given the significant difference 
			 in heliospheric latitude between the two. \\}
\end{table}

In addition to accurately modeling the cometary emissions, we compare our SW compositions results to contemporaneous composition ratios provided by \textit{ACE}. A comparison of our model results to \textit{ACE}, shown in Table \ref{ACE}, demonstrate an agreement within uncertainty for all observations. These results provide an additional, and crucial, confirmation for the physical accuracy of our modeling technique. It also leads us to consider using our analysis and modeling of cometary X-ray  spectra in the future as a remote diagnostic tool for SW composition.


\section{\label{Discussion}Discussion}

\subsection{\label{HRC_sect}X-Ray Emission Morphology}

When analyzing cometary X-ray images, it is important to remember that the overall emission morphology is determined by SW interaction with the cometary atmosphere as the majority of the emitted intensity is due to CX. In the collisionally thick case for an active comet, we expect a paraboloid with the comet at the focus where the magnitude of the semimajor axis is dependent on the atmospheric density \citep{Haberli1997, Wegmann2004, Lisse2005, Wegmann2005}. In comparison, the X-ray emission structure is determined by the distribution of gas in the coma for the collisionally thin case. For such a situation, we expect to see regions of enhanced X-ray emission in regions of higher cometary particle density, such as those found along jet structures \citep{Lisse2013}.

\begin{figure*}[t]
	\epsscale{1.15}
	\plotone{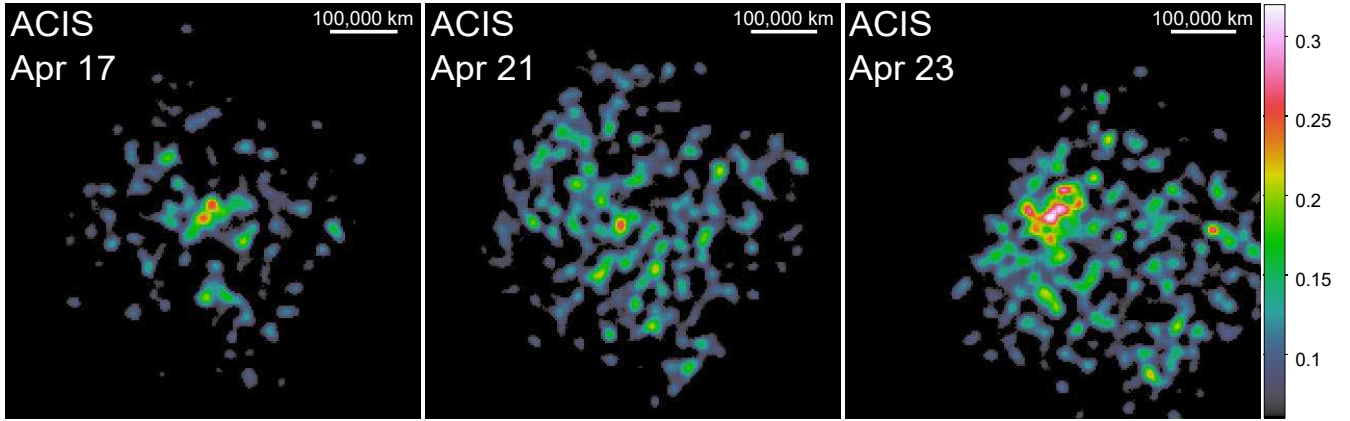} 
	\caption{\textit{Chandra} /ACIS-S observations of comet PanSTARRS. 
	The images are binned to include all 0.3--1.1 keV photon events, exposure corrected, 
	 on the same linear scale, and smoothed with a 5 $\times$ 5 pixel Gaussian filter. 
	Our results show fluctuations in X-ray emission intensity, 
	and the overall morphology is highly non-uniform.}
\label{PS_images}
\end{figure*}

We present the ACIS observation images of comet PanSTARRS in Figure \ref{PS_images}. All images have been corrected for differences in exposure time and are normalized to the same linear scale. These images show a constant intensity in X-ray emissions in all the observations, as we expect given the constant cometary dust/gas emission rates and SW conditions observed at the time of our observations. We also find that the overall morphology is highly non-uniform, and so we conclude that PanSTARRS was collisionally thin during its observations. This is likely a result of the high dust density present in the cometary atmosphere as dust particles are significantly less efficient in CX X-ray production than molecular gas \citep{Djuric2005,Wolk2009,Lisse2013}.

\begin{figure*}[t]
	\epsscale{1.15}
	\plotone{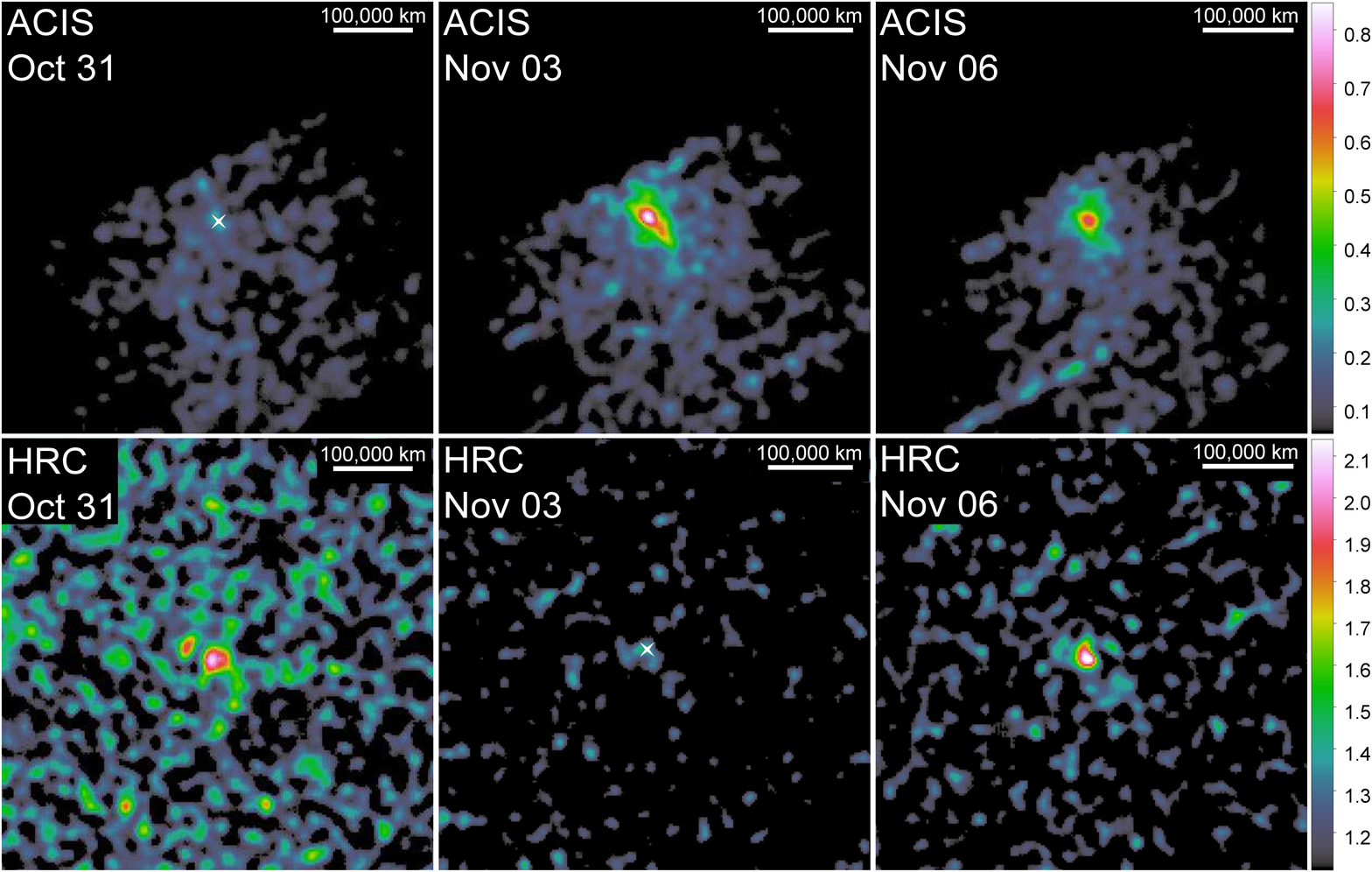} 
	\caption{\textit{Chandra}/ACIS-S and HRC-I observations of comet ISON. Each set of images
	(either ACIS or HRC) are exposure corrected, shown on a linear scale, 
	and smoothed with a 5 $\times$ 5 pixel Gaussian filter. ACIS images are also binned to include
	all 0.3--2.0 keV photon events. The images show a ``see-saw" effect between the soft X-ray HRC 
	observations and hard X-ray ACIS observations where an increase of intensity in HRC correlates 
	to a decrease in ACIS, and vice versa. This result also correlates to fluctuations in SW 
	speed between October 31 and November 03 as seen via \textit{ACE}.}	
\label{ISON_images}
\end{figure*}

The ISON observations were unique as it was the first time HRC observations of a comet were performed in conjunction with ACIS observations. The extracted ACIS and HRC images for the three ISON observations are shown in Figure \ref{ISON_images}. Each image set, either from ACIS or HRC, has been corrected for differences in exposure time and has been normalized to the same linear scale. The ACIS images demonstrate the expected paraboloid morphology of a collisionally thick case, while the HRC observations depict a more non-uniform emission typical of a collisionally thin case. Given that HRC is more sensitive to soft X-rays than ACIS, this result may indicate that the soft X-ray emissions are due to CX emissions from lighter SW ions with smaller cross sections, such as He$^{2+}$, than the SW ions that emit hard X-rays, such as C$^{6+}$ and O$^{8+}$. The reduction in cross section will allow deeper penetration into the cometary atmosphere, and it may be substantial enough to generate difference in these image sets. It is also possible that the soft X-ray emissions from ISON are from a different emission mechanism that would not produce the same morphology, such as scattering or fluorescence. 

The ISON image sets also demonstrate significant fluctuation in the cometary emission intensity over time and a ``see-saw" in intensity between the soft X-ray HRC observations and the hard X-ray ACIS observations, most notably seen on the October 31 and November 03 visits. These intensity fluctuations correlate with increases in SW speed as documented by \textit{ACE}, where the maximum SW speed was recorded November 03. Such an association between SW and cometary emission intensity is predicted by our CX model  as SW speed fluctuations indicate fluctuations in SW ion freeze-in temperatures \citep{Bochsler2007}. Such temperature changes will shift the SW charge state distribution, producing a varying average cometary emission energy based on our normalized photon emission yield function $P_{k,l}^{(j)}(\hbar \omega_{j})$. As we see a similar shifting of the average cometary emission energy, we therefore assert that CX emissions are the dominant cometary emission mechanism in the soft X-ray region, a fact that will become important in our discussion in Section 4.4.

\subsection{PanSTARRS Spectral Analysis}

Prior to the observation of comet PanSTARRS, there was much speculation if its high dust-to-gas ratio would significantly affect its X-ray CX emission intensity as it is more favorable to produce Auger electrons instead of X-rays when undergoing CX with dust particles \citep{Djuric2005, Wolk2009,Lisse2013}. We therefore make sure to note any X-ray spectral irregularities within our results and, if so, their possibility of being due to dust particles. 

Utilizing our CX model, we are able to successfully characterize PanSTARRS' spectrum without making any adjustments to our CX scenario. We find a unique solution for the emission spectrum that fits well to the observations up to 1.0 keV. Above 1.0 keV, the uncertainty in the observations becomes too great to distinguish between noise and emission peaks. Analysis of SW composition through the use of our model shows a lower than average amount of highly charged ions, such as O$^{8+}$ and Ne$^{9+}$, with an increase in their lower energy variants, like O$^{6+}$ and Ne$^{8+}$. This result agrees with our previous assessment that PanSTARRS was observed at fast, polar SW. Beyond this irregularity, PanSTARRS' spectrum possesses no additional traits that would classify it different from any other comet X-ray spectrum.  

Although we cannot infer from our spectral analysis how PanSTARRS' large dust quantities may have impacted other emissions mechanisms present within the cometary spectra, our results indicate that it had little to no observable impact on the comet's CX X-ray emissions. As such, any differences present are more likely attributed to the SW flux density and ionization state at the time of observation.

\subsection{ISON Spectral Analysis}

Despite being one of the brightest comets in recent years, the \textit{Chandra} observations we analyzed were taken slightly prior to ISON's drastic increase in gas production rate starting on 2013 November 13. Fluctuations in SW speeds, as confirmed by \textit{ACE}, and several M-class solar flares, as reported by \textit{GOES}, were also observed during ISON's \textit{Chandra} visits. These highly volatile SW conditions may significantly impact ISON's emission spectra. 

Using the ACIS observations and applying the same method as done for PanSTARRS, we are able to model ISON's spectrum as CX below 1 keV and extract SW composition ratios. ISON's ratios confirm the above average SW speed with an overabundance of highly charged SW ions, like O$^{8+}$ and Ne$^{9+}$, that produces a distinct plateau in the spectrum from 0.75--1.00 keV. The O$^{7+}$ ratio is twice that seen from PanSTARRS, best visualized via the emission peak at 0.6 keV. 

In addition to these results, ISON exhibits some possible peak structures in its emission spectrum at energies above 1 keV: one peak at 1.35 keV and another at 1.85 keV, as seen in Figure \ref{comets}. Such peaks have been previously seen in \textit{Chandra}'s observations of Comet 153P (Ikeya-Zhang) \citep{Ewing2013}, another comet viewed during volatile SW conditions. Our model is presently unable to accurately calculate theoretical CX emissions in this energy range due to the lack of information about the presence of such highly charged ions in the solar wind plasma, but we may comment on the possible emission candidates. 

Comparison to atomic emission line tables from \textit{NIST} indicate that the most probable ions for each emission is Mg XI \mbox{(1s$^2$ {}$^{1}$S--1s2p {}$^{1,3}$P)} for 1.35 keV and either Si XIII \mbox{(1s$^2$ {}$^{1}$S--1s2p {}$^{1,3}$P)} or Mg XII \mbox{(1s {}$^{2}$S--4p {}$^{2}$P)} for 1.85 keV \citep{NIST2015}. However, it is unclear if these peaks could be a result from CX as these exotic candidates have not been detected via in situ observations of SW ion composition \citep{vonSteiger2000,Lepri2013}. Furthermore, theoretical models describing the charge abundance of heavy SW ions predict an extremely low probability of finding these ions because of the inability to reach such high freezing-in temperatures in regular SW and coronal mass ejections \citep{Bochsler2007}. On the other hand, these spectral lines are clearly presented in the spectra of the solar X-ray flares as well as in a regular X-ray emission from the Sun \citep{McKenzie1985, Dere1997, Landi2013}. It therefore may be possible these peaks are due to a different mechanism whose emissions are increased by solar flare activity, such as scattering of solar X-rays \citep{Krasnopolsky1997,Snios2014}. At present, we cannot conclude what is the primary source of these exotic emissions detected from ISON as further analysis of ISON's spectrum with a revised CX model, and possibly also a scattering emission model, is required. 

\subsection{\label{soft_xray}Potential Soft X-Ray Emissions from ACIS}

\begin{figure}[t]
	\epsscale{1.15}
	\plotone{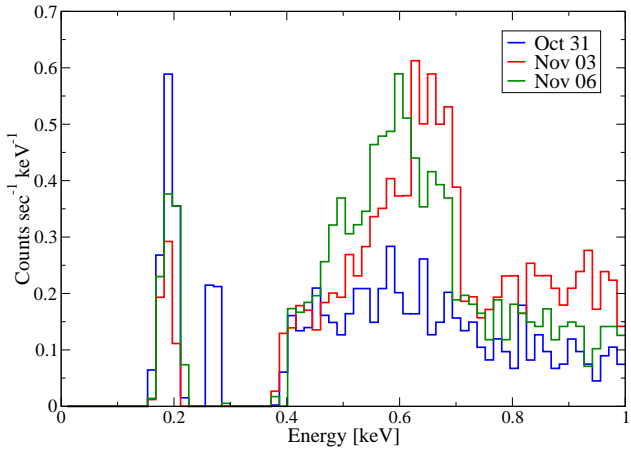} 
	\caption{ACIS spectral intensity for each observation of comet ISON. 
		All detector contamination from the carbon K-shell at 0.284 keV has been 
		removed from the spectra. The observed spectral feature at 0.2 keV has 
		the the same fluctuation in intensity that is observed from the HRC observations
		and exceeds the average spectral intensity uncertainty in this region, 
		and so we conclude that this feature is physical. Possible origins of this spectral 
		feature are discussed in Section 4.4.}
	\label{ACIS_ISON}
\end{figure}

While examining the ACIS spectra from ISON, we found a peak-like feature located at 0.2 keV. This feature was also found in the PanSTARRS observations, but with a lower relative spectral intensity. After removing the portion of the spectra caused by the carbon K-shell detector contamination at 0.284 keV, we plot the resulting emission spectra for each of the three ISON visits in Figure \ref{ACIS_ISON}. Our plots show a soft X-ray region at 0.2 keV that the detector is sensitive to after our corrections, even showing fluctuations that agree with the soft X-ray emission fluctuations detected by HRC (see Figure \ref{ISON_images}). The overall shape of this feature is likely due to the ACIS effective area function abruptly decaying toward zero at 0.18 keV and is not due to any specific emission line. Also, the fluctuations between the visits exceed the spectral intensity uncertainty in this region, which is 0.08 counts s$^{-1}$ keV$^{-1}$, and so we believe these features to be physical.  

\begin{figure*}[t]
	\epsscale{1.17}
	\plottwo{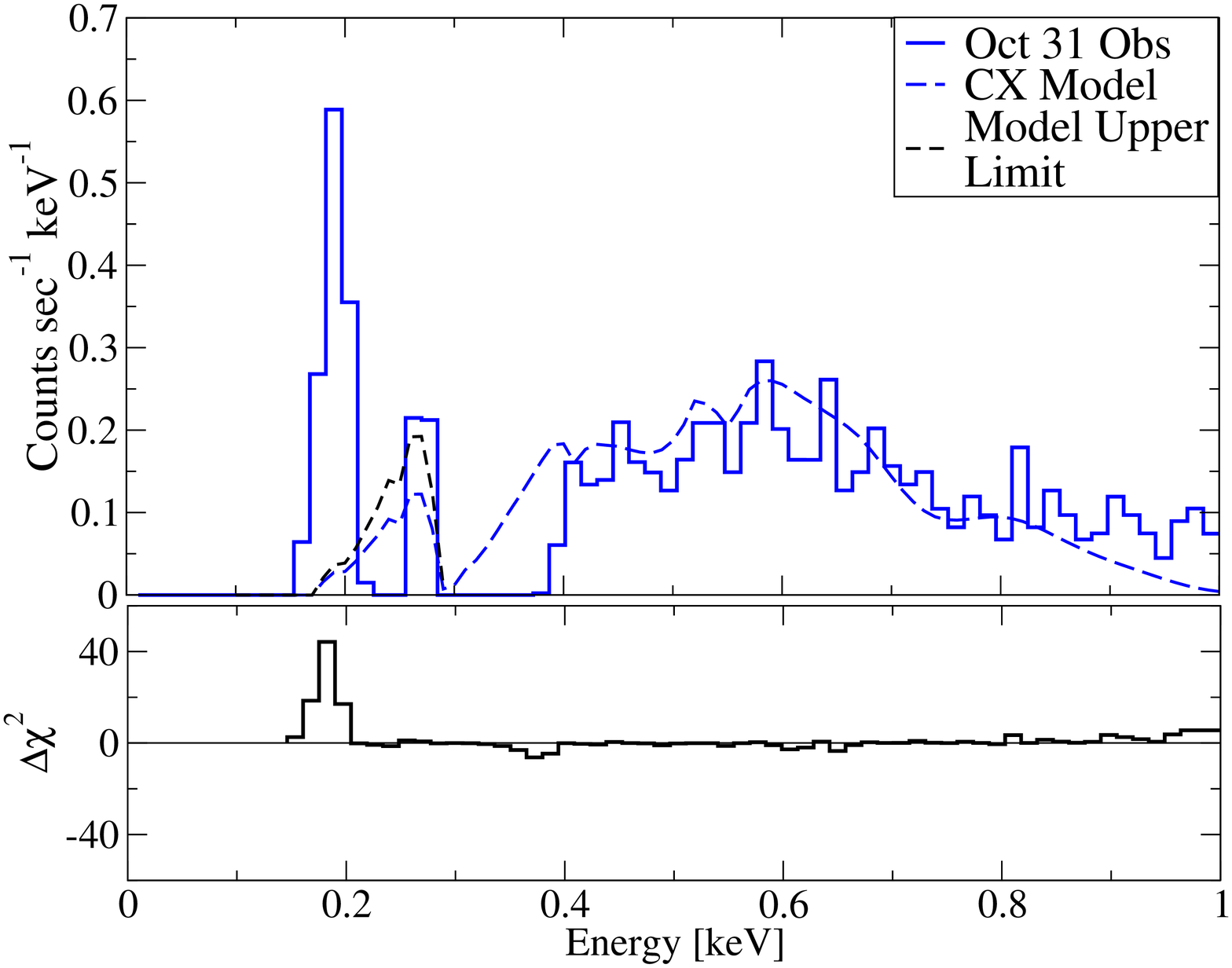}{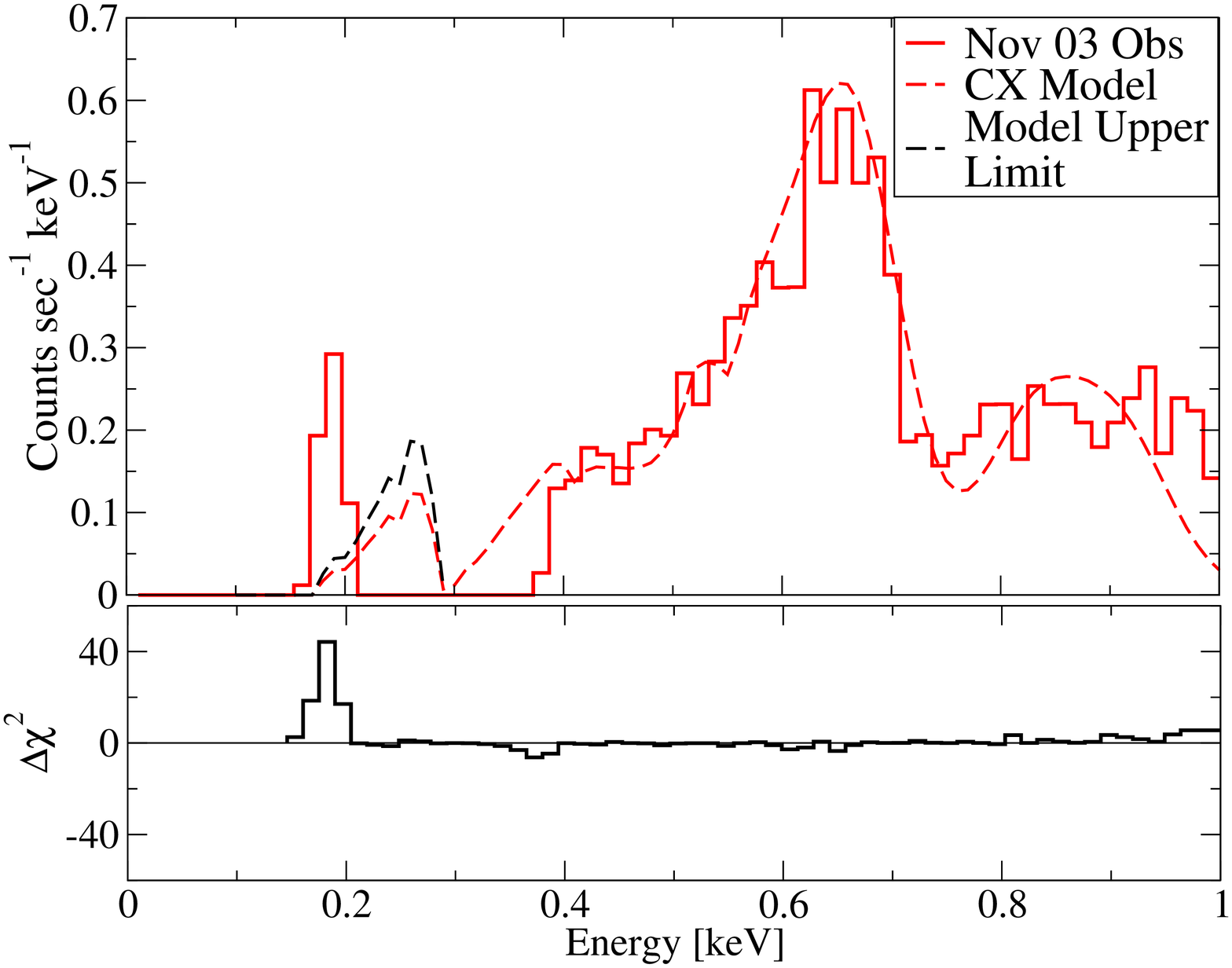}
	\epsscale{0.57}
	\plotone{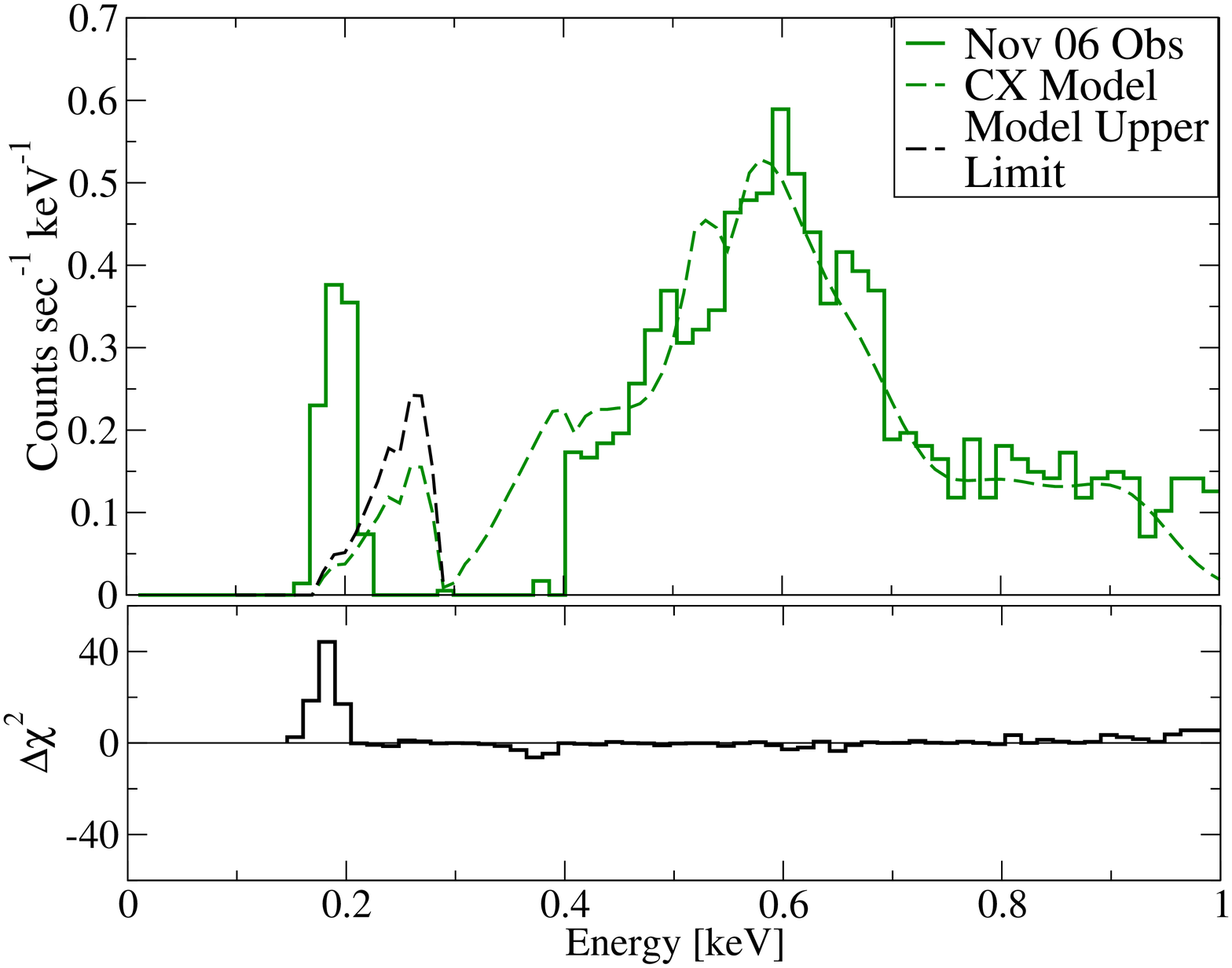}
	\caption{ACIS spectral intensity for each observation of comet ISON (solid lines)
		and its respective modeled CX using average SW compositions (dotted colored lines). 
		All detector contamination from the carbon K-shell at  0.284 keV has 
		been removed from the spectra. Despite the excellent agreement above 0.4 keV, 
		our model fails to match the shape and intensity of the soft X-ray spectral feature. 
		We also calculate an upper limit to soft X-ray CX emissions by accounting for 
		sequential CX events and assuming that all SW ions are neutralized through interaction with the 
		cometary atmosphere (dotted black lines), and our results show these additions to be
		insufficient to equal the observed soft X-ray intensities.
		We therefore believe these soft X-ray features to be CX from an unaccounted SW ion 
		(such as He$^{2+}$), detector contamination, or a combination of these options. }
	\label{ACIS_ISON_MODEL}
\end{figure*}

Based on our HRC results from Section 4.1, we assume CX emissions to be the most likely cause of this feature. We therefore extend our CX model down to this soft X-ray region and plot the results with the observational data. See the dotted lines in Figure \ref{ACIS_ISON_MODEL} for our predicted CX model for each observation. Calculation of a unique solution of SW ratios required to produce such intensities is not possible due to the abundance of over 200 unique lines from over 15 different SW ion types that fall within ACIS' resolution of this soft X-ray feature. We therefore choose to leave our model at average SW abundances in this region. We note that fixing these parameters produces no difference in the spectral fit over the 0.3--1.0 keV energy range.

Our results show that our average CX model is not capable of producing the necessary intensity to match the observations for either the ISON emissions or the PanSTARRS emissions, which are not shown. Furthermore, the SW abundances that our model would demand to match these features in intensity far exceed their physical boundaries, with most abundances requiring an increase by an order of magnitude. Although such exotic SW compositions are not impossible given its constantly fluctuating nature, the consistent presence of these soft X-ray features during both fast and slow SW indicate these should be generated under average SW conditions.

Although our current CX model does not agree with the soft X-ray intensities detected, we only consider a single electron capture event per incoming SW ion. Sequential capture events may occur for an ion if the cometary atmosphere is collisionally thick, increasing the amount of soft X-rays emitted from the system as the ion charge state decreases (O$^{8+}$$\rightarrow$ O$^{7+}$$\rightarrow$ O$^{6+}$$\rightarrow$ ... ). We therefore modify our CX model to include these sequential capture events per ion as it may solve our soft X-ray intensity deficit.

For our analysis, we calculate an upper limit on the increase to soft X-ray CX emissions from sequential capture events by assuming all SW ions are neutralized through interaction with the cometary atmosphere. Our results are presented in Figure \ref{ACIS_ISON_MODEL}, and they show that the additional CX events are not sufficient to equal the observed soft X-ray intensities. We find that the upper limit of CX emissions only increases the total soft X-ray intensity $\sim$50\%, which is not enough to account for the factors of three to six between the model and the observations. Furthermore, we stress that the actual rate of sequential capture events present in these cometary systems is lower than this upper limit, so the actual emission intensities will reside between our model and the upper limit. We therefore find it unlikely that sequential CX events could account for these soft X-ray features.

As we are confident that both our CX model's resulting SW composition and photon yield emission rates are accurate, we therefore consider two possible explanations for the soft X-ray discrepancy: 

\begin{enumerate} 

\item Since the CX model does not match the observational intensities, it is possible we lack the SW ion type required to produce this feature. He$^{2+}$ CX emissions, currently not included in our analysis,  would be detectable in this soft X-ray region due to the low resolution of ACIS, and its high abundance may provide the required order-of-magnitude increase in intensity \citep{Kharchenko2001,Bodewits2004}. Such SW ions would also have deeper penetration in the cometary atmosphere, which might also explain the collisionally thin appearance of the HRC morphology discussed in Section 4.1. Future iterations of our model should include this ion and compare the modified results to the soft X-ray emissions from ACIS. 

\item The soft X-ray feature may be a result of previously undocumented detector contamination or degradation that sharply decays below 0.2 keV, producing a peak in observed spectrum. Examination of similar comets observed at different stages of ACIS' lifetime would show if such a soft X-ray feature is always present, indicating cometary origins, or if this feature has manifested itself over time, indicating a detector issue. 

\end{enumerate} 

The required analysis for each of these possibilities is beyond the scope of this article, but we believe that any future work on these soft X-ray features from ACIS should provide a thorough analysis of each possible explanation to determine the cause of these unique findings.


\section{\label{Conclusions}Conclusions}

In summary, we have used Chandra to study two very different Oort Cloud comets, the gas-rich C/2012 S1 ISON and the dust-rich C/2011 L4 PanSTARRS. Both comets were observed within 1 AU heliocentric distance of the Sun, when they were active. The observed X-ray morphologies were dramatically different, however, with ISON displaying an extended, well-developed X-ray coma and PanSTARRS producing an unformed X-ray haze. The two comets also experienced markedly different SW conditions, with ISON impacting an excited wind, while PanSTARRS traveled through fast SW. 

We developed an updated CX emission model that includes large amounts of ion spectral lines induced in CX collisions and simplifies input variables while improving the physical accuracy in comparison to previous models. Our model was used to analyze \textit{Chandra} observations of comets ISON and PanSTARRS, and we found strong emissions induced in CX collisions of SW ions normally present within cometary emissions (C$^{5+}$, C$^{6+}$, N$^{5+}$, N$^{6+}$, N$^{7+}$, O$^{6+}$, O$^{7+}$, O$^{8+}$,  Ne$^{8+}$, Ne$^{9+}$) from both comets. Analysis of ISON spectra shows higher concentrations of O$^{8+}$ and Ne$^{9+}$ than PanSTARRS, indicating higher SW ion freeze-in temperature during its observations.

Analysis of ISON's spectrum also shows high-energy spectral features above 1 keV. To clarify the physical origin of these cometary ``hard" X-rays, we intend in the future to include CX emissions from exotic SW ions, such as Mg$^{11+}$ and Si$^{13+}$, which will extend our model beyond 1 keV. Analysis of high-energy spectral features will allow us to predict the total ratio of exotic SW ions and to discuss whether those quantities could be observable using current tools. We will also include emission contributions from scattering and fluorescence of energetic solar X-rays, especially during solar X-ray flares events, and X-ray emissions of non-thermal energetic electrons due to electronic impact or bremsstrahlung mechanisms. Accurate investigations of the spectral morphology, which are different for each mechanism, will also be performed. Such a discussion would establish a hierarchy of potential contributing mechanisms in cometary X-ray spectra above 1 keV and provide insight on the origin of the observed high energy spectral features.

Beyond successfully analyzing comets ISON and PanSTARRS, we also demonstrated our model's potential use as a SW ion composition analyzer. Our composition results agree well with other SW composition tools available, such as \textit{ACE}, while also calculating unique composition ratios not available through these other tools, like Ne$^{8+}$, Ne$^{9+}$, and Mg$^{10+}$. With further development of CX X-ray modeling, such an application would be possible for any CX emissions, not just those from comets. Our model also simplifies the variable inputs and provides an additional information on SW composition. We therefore intend to use such a model for all future CX analyses of cometary and planetary X-ray emissions as well as for investigations of CX X-rays induced in interaction between the SW plasma and interstellar gas. 

In addition to our modeling results, we found the possibility of soft X-ray emissions around 0.2 keV detected from both comets ISON and PanSTARRS via ACIS. These soft X-ray features fluctuate similarly to those observed from the HRC observations and exceed the average spectral intensity uncertainty, leading us to believe these features to be cometary CX in origin. We extended our CX model to this soft X-ray region to compare, only to find our results lower in intensity than the observations by an order of magnitude. We also revise our model to include sequential CX capture events as it will incease soft X-ray intensities, but we find that even the inclusion of more capture events is not sufficient to match the observed intensities. Based on our confidence in the model from its previous results, we believe this discrepancy to be a result of either a lack of SW ion types that produce significant emissions in the soft X-ray region (such as He$^+$), detector contamination or degradation, or a combination of these possibilities. Investigations of these soft X-ray features should carefully explore each explanation as confirmation of these features as physical emissions would open new opportunities in understanding cometary emission processes via \textit{Chandra}. 
 

\begin{acknowledgements}
	We would like to acknowledge National Oceanic and Atmospheric Administration for their \textit{Advanced Composition Explorer} data. 
	The work of B.S. and V.K. on this project has been supported by NASA grant NNX10AB88G. 
	The work of C.L. has been supported by Chandra grant GO4-15001A.	
\end{acknowledgements}


\bibliographystyle{aa}
\bibliography{all_data}

\end{document}